\documentclass[aps,prl,twocolumn,superscriptaddress,longbibliography]{revtex4-1}
\pdfoutput=1
\usepackage{bm}
\usepackage{graphicx}
\usepackage{booktabs}
\usepackage{slashed}
\usepackage{hyperref}
\usepackage{amssymb,amsmath,amsfonts}
\usepackage{color}
\usepackage[utf8]{inputenc}
\usepackage[caption=false]{subfig}
\newcommand{\pol}{{\mathbb P}}
\newcommand{\pot}{ {\mathcal V} }
\newcommand{\pth}{{\mathcal P}}
\newcommand{\wilson}{{\mathcal W}}
\newcommand{\partition}{{\mathcal Z}}

\newcommand{\curve}{{\mathcal C}}
\newcommand{\vx}{\vec{x}}
\newcommand{\vy}{\vec{y}}
\newcommand{\hamil}{{\mathcal H}}
\newcommand{\rep}{{\mathcal R}}

\newcommand{\tlrep}{\widetilde{{\mathcal R}}}

\newcommand{\chg}{{\mathcal Q}}
\begin{document}

\title{Wilson loops in the Hamiltonian formalism}

\author{Robert D. Pisarski}
\affiliation{Department of Physics, Brookhaven National Laboratory, Upton, NY 11973}

\begin{abstract}
  In a gauge theory, the gauge invariant Hilbert space is unchanged by the coupling to arbitrary local
  operators.  In the presence of Wilson loops, though, the physical Hilbert space must be enlarged
  by adding test electric charges along the loop.
  I discuss how at nonzero temperature Polyakov loops are naturally related to the propagator of a test charge.
  't Hooft loops represent the propagation of a test magnetic charge, and so
  do not alter the physical Hilbert space.
\end{abstract}

\maketitle

Since the seminal work of Wegner, Wilson, and Creutz, simulations of lattice gauge theories
using Monte Carlo methods on classical computers has given us invaluable
information about Quantum ChromoDynamics (QCD).
For the phase diagram at a nonzero temperature $T$ and quark chemical
potential $\mu$, at $\mu = 0$ the order parameter for chiral symmetry exhibits crossover behavior 
at a temperature of $T_\chi \approx 156 \pm 1.5$~MeV
\cite{HotQCD:2018pds,Borsanyi:2020fev,Guenther:2022wcr}
\footnote{This accuracy should not obscure the fact that the crossover in QCD is rather broad, over {\it tens} of MeV;
the lattice precisely measures the maximum of a slowly varying function \cite{Guenther:2022wcr}.}.
These methods have been extended to quark chemical potentials less than the temperature,
$\mu \leq T_\chi$ \cite{Borsanyi:2020fev,Guenther:2022wcr,Bazavov:2020bjn,Bollweg:2021vqf,Borsanyi:2021sxv,Ratti:2021ubw}.

This leaves many quantities of direct experimental significance which have not yet been computed.  For
example, the diffusion coefficient for a heavy quark has been computed in the $SU(3)/Z(3)$ gauge theory, without
dynamical quarks \cite{Brambilla:2020siz,Altenkort:2020fgs}.  The computation of other transport coefficients,
notably the shear and bulk viscosities with dynamical quarks in QCD,
is conceivable with much larger classical computers
\footnote{S. Mukherjee, private comment.}.

Many other quantities, such as correlation functions in real time, and the properties of cold, dense QCD,
are only possible with quantum computers.  While large scale quantum computers with logical qubits
lie well in the future, it is useful to consider the questions of principle which are unique to a gauge theory.

A classical computer deals with the Lagrangian.  If the chemical potential vanishes, at
any temperature the action is real, and sophisticated techniques, including those for nearly massless quarks, have
been developed.  While in principle many states contribute to the partition function, the Metropolis
algorithm automatically selects the most important.  The difficulty is the sign problem:
at nonzero chemical potential the action is no longer real, and standard techniques fail.

In contrast, for a quantum computer it is best to deal with the Hamiltonian.  
The partition function is
\begin{equation}
  \partition = \sum {\rm e}^{- {\mathcal H}/T - \mu {\mathcal N} } \; ,
\end{equation}
where ${\mathcal H}$ is the Hamiltonian, ${\mathcal N}$ the quark number density,
and $\sum$ is the sum over all states.  Because everything is real,
there is no sign problem when $\mu \neq 0$.  The difficulty
is that exponentially many states contribute, and even for states near the ground state, it is not obvious how
to choose the most important.
Strategies to solve this have been developed in condensed matter systems,
and include the density matrix renormalization group, matrix product states, and
projected entangled pair states \cite{Cirac:2020obd,Shachar:2021vbu}.

While some generalization of these methods will be essential in QCD, the purpose of this paper is to
make an elementary point about how the Hamiltonian form of a gauge theory changes in the presence of Wilson loops.

For a theory without gauge fields, the most general correlation functions are given by
adding sources for arbitrary local operators to the Lagrangian.  Multiple insertions of composite operators induce
new counterterms to the theory, but this is standard
\cite{Zinn-Justin:2002ecy,Skokov:2010sf,Pisarski:2016ixt}, and going from the Lagrangian
to the Hamiltonian formalism is direct.

With gauge fields, however, there are gauge invariant non-local operators, such as the Wilson loop,
\begin{equation}
  \wilson_\curve = {\rm tr} \, \pth \exp\left( i g \int_\curve A_\mu \, dx^\mu \right) \; .
  \label{eq:wilson_loop}
\end{equation}
Here $g$ is the gauge coupling, $A_\mu$ is the vector potential for the gauge field,
$\pth$ denotes path ordering along a closed curve $\curve$, and the trace is over color.

My basic point is simple.  Dynamical quarks contribute to Gauss's law at each
point in space.  If the quark mass is sent to infinity, so all that remains
is one test quark propagating along $\curve$, then Gauss's law must include the contribution of
that test quark \cite{Gervais:1978kn,Echevarria:2020wct}.  I show in this paper how the sum over
states of the test quark generates the Wilson loop, $\wilson_\curve$.

I begin with the Lagrangian formalism, where the analysis is transparent, and use that to proceed 
to the Hamiltonian form.  While I consider
systems which are independent of time, with the Hamiltonian formalism it is possible to
perturb a gauge theory with a gauge-invariant, time dependent source, and then measure the evolution 
of Wilson loops in time.  This validates computing the ``holonomous'' potential
for the eigenvalues of the thermal Wilson line
\cite{Gross:1980br,Weiss:1980rj,Bhattacharya:1990hk,Belyaev:1991gh,Bhattacharya:1992qb,Gocksch:1993iy,KorthalsAltes:1993ca,KorthalsAltes:1994be,KorthalsAltes:1996xp,KorthalsAltes:1999cp,Giovannangeli:2001bh,Giovannangeli:2002uv,Giovannangeli:2004sg,Dumitru:2013xna,Guo:2014zra,Nishimura:2017crr,Guo:2018scp,KorthalsAltes:2019yih,KorthalsAltes:2020ryu,Hidaka:2020vna,Guo:2020jvc},
and using it to construct effective theories for the deconfining \cite{Polyakov:1978vu,Susskind:1979up} and chiral
\cite{Pisarski:1983ms} phase transitions 
\cite{Pisarski:2000eq,Dumitru:2000in,Dumitru:2001xa,Dumitru:2003hp,Dumitru:2004gd,Dumitru:2005ng,Oswald:2005vr,Pisarski:2006hz,Hidaka:2008dr,Hidaka:2009hs,Hidaka:2009xh,Hidaka:2009ma,Dumitru:2010mj,Dumitru:2012fw,Kashiwa:2012wa,Pisarski:2012bj,Kashiwa:2013rm,Lin:2013qu,Bicudo:2013yza,Smith:2013msa,Lin:2013efa,Bicudo:2014cra,Gale:2014dfa,Hidaka:2015ima,Pisarski:2016ukx,Pisarski:2016ixt}.
Incidentally, it alleviates some concerns
\cite{Belyaev:1991np,Chen:1992sa,Kogan:1993bz,Smilga:1993vb,Hansson:1994ep,Kiskis:1995iq,Smilga:1996cm,Korthals-Altes:1999cqo,Korthals-Altes:2000tia,deForcrand:2000fi,Cohen:2022gyi}
about holonomous potentials

A Wilson loop alters the Hilbert space because it represents the propagation of a test electric charge.
In contrast, the 't Hooft loop
\cite{tHooft:1979rtg,tHooft:1980kjq,Korthals-Altes:1999cqo,Korthals-Altes:2000tia,deForcrand:2000fi,deForcrand:2001nd,deForcrand:2005pb,Reinhardt:2002mb,Reinhardt:2007wh}
represents the propagation of a test magnetic charge, and so doesn't modify
the Hilbert space.  I also show how to compute 't Hooft loops in the Hamiltonian formalism.

{\bf Lagrangian formalism:}
Consider a quark of mass $M$ as $M \rightarrow \infty$.   Then we can neglect the spin of the quark,
as spin-dependent effects are uniformly suppressed as $\sim 1/M$.  Similarly, we can consider either a
quark, propagating forward in time, or an anti-quark, propagating backwards in time.
In either case, since it is too heavy to move, the quark (or anti-quark), just 
sits at some point in space
\footnote{Boosting to a moving frame gives a test quark moving at constant velocity.}.
The gauge invariant effective Lagrangian is then
\begin{equation}
  {\mathcal L} = \psi^\dagger \,  D_0 \, \psi \; .
  \label{eq:lag_inf_mass}
  \end{equation}
I assume that $\psi$ lies in the fundamental representation,  
where the covariant derivative $D_0 = \partial_0 - i g A_0$.  Later I generalize to arbitrary representations.

I introduce the Wilson line, running from a point $x$ in space-time to $y$:
\begin{equation}
  {\bf L}(y;x) = \pth \exp\left( i g \int^{y}_{x} A_\mu(z) \, dz^\mu \right) \; ,
  \label{eq:def_line}
\end{equation}
uniformly taking a straight line path between the two.  Regardless of the path, the Wilson
line transforms homogeneously under a gauge transformation $\Omega(x)$,
\begin{equation}
  {\bf L}(y;x) \rightarrow \Omega^\dagger(y) \, {\bf L}(y;x) \, \Omega(x) \; .
\end{equation}
Since $D_0 \, {\bf L}(\vx; t',t) = 0$, the propagator for a test quark sitting at a point
$\vx$ is proportional to ${\bf L}(\vx; t',t) = {\bf L}(\vx,t';\vx,t)$, using an obvious abbreviated notation.

A test meson is constructed by putting a test quark at one point, $(\vx,0)$, and tying it with a 
spatial Wilson line to a test anti-quark at $(\vec{0},0)$.  
A rectangular Wilson loop represents the propagation of this test meson up in time, until
it is annhilated by a test anti-meson.

At nonzero temperature it is possible to construct a gauge invariant operator for a single test quark.
In most gauges the gauge fields can be taken to periodic in imaginary time,
$A_\mu(\vx,1/T) = + A_\mu(\vx,0) $.
The Polyakov loop is the trace of the full thermal Wilson line, which runs from $\tau: 0 \rightarrow 1/T$,
\begin{equation}
  \pol(\vx) = {\rm tr} \; {\bf L}(\vx;1/T,0)  \; .
  \label{eq:polyakov_loop}
\end{equation}
This is invariant under strictly periodic gauge transformations.
In a gauge theory without dynamical quarks, though, the gauge symmetry is $SU(N)/Z(N)$,
and it is also necessary to consider global $Z(N)$ transformations.
These form the center of the gauge group, $\omega_j = {\rm e}^{2 \pi i j/N} \, {\bf 1}$, $j= 1\ldots N$,
where the $\omega_j$ commute with all elements of the group.
Then more general gauge transformations are possible, which are periodic only up to a factor of $\omega_j$,
\begin{equation}
  \Omega(\vx,1/T) = \omega_j \, \Omega(\vx,0) \; ,
  \label{eq:twist}
\end{equation}
Since the $\omega_j$ commute with all group elements, the gauge fields remain strictly periodic.
The Polyakov loop, though, transforms linearly,
\begin{equation}
  \pol(\vx) \rightarrow \omega_j \, \pol(\vx) \; .
\end{equation}
The spatial average of the vacuum expectation value, $\int d^3 x \, \langle \pol(\vx) \rangle$,
is an order parameter for the spontaneous breaking of a global, one-form \cite{Gaiotto:2014kfa} $Z(N)$ symmetry
\cite{Polyakov:1978vu,Susskind:1979up}.  Without dynamical quarks, this is an exact symmetry in the confined
phase, which is spontaneously broken in the deconfining phase.  With dynamical quarks the Polaykov loop
is still a gauge invariant operator, but the global $Z(N)$ symmetry is lost.

Averaging over space, the expectation value of the Polyakov loop is
\begin{equation}
\langle \pol \rangle  = \int \frac{d^3x}{V} \; \int {\cal D} A_\mu \; {\rm e}^{- \, {\cal S}(A_\mu)}\;
    \langle \pol(\vx) \rangle \; .
  \label{eq:free_energy}
\end{equation}
This is the path integral over the gauge field, with action ${\cal S}(A_\mu)$, 
averaged over the spatial volume $V$.
This average is not a free energy, as sometimes stated
\cite{Belyaev:1991np,Chen:1992sa,Kogan:1993bz,Smilga:1993vb,Hansson:1994ep,Kiskis:1995iq,Smilga:1996cm,Korthals-Altes:1999cqo,Korthals-Altes:2000tia,deForcrand:2000fi}.
Assume that $\langle \pol \rangle = {\rm e}^{- {\mathcal F}_\infty/T}$.
If ${\cal F}_\infty$ were a free energy, then it must be real, with an exponential which is real and positive.
For two colors, however, $\pol(\vx)$ can be negative, while for three or more colors, it is complex.

As the color trace of a propagator, though, there is nothing amiss if
the Polyakov loop is either negative or complex.
Consequently, the Polyakov loop, or more properly the eigenvalues of the thermal Wilson line
\cite{Gross:1980br,Weiss:1980rj,Bhattacharya:1990hk,Belyaev:1991gh,Bhattacharya:1992qb,Gocksch:1993iy,KorthalsAltes:1993ca,KorthalsAltes:1994be,KorthalsAltes:1996xp,KorthalsAltes:1999cp,Giovannangeli:2001bh,Giovannangeli:2002uv,Giovannangeli:2004sg,Dumitru:2013xna,Guo:2014zra,Nishimura:2017crr,Guo:2018scp,KorthalsAltes:2019yih,KorthalsAltes:2020ryu,Hidaka:2020vna,Guo:2020jvc},
are not an artifact of Euclidean spacetime
\cite{Belyaev:1991np,Chen:1992sa,Kogan:1993bz,Smilga:1993vb,Hansson:1994ep,Kiskis:1995iq,Smilga:1996cm,Korthals-Altes:1999cqo,Korthals-Altes:2000tia,deForcrand:2000fi,Cohen:2022gyi}, but perfectly sensible physical quantities.

At nonzero temperature, the potential between a test quark and anti-quark is given by the two point function of
Polyakov loops,
\begin{equation}
{\rm e}^{-\pot_\infty(\vx)/T} = \langle \pol^\dagger(\vx) \pol(\vec{0})\rangle -|\langle \pol \rangle|^2 \; .
\end{equation}
Another gauge invariant quantity is the thermal Wilson loop,
\begin{equation}
  \wilson_T = {\rm tr} \, {\bf L}(\vec{0};1/T,0)  {\bf L}(\vx,\vec{0};1/T)  {\bf L}(\vx;0,1/T)  {\bf L}(\vec{0},\vx;0) \; .
\end{equation}
Because of the spatial Wilson lines from $\vec{0}$ to $\vx$ at $\tau = 0$ and $1/T$,
contributions which don't appear in $\pot_\infty(\vx)$ enter \cite{Laine:2006ns}.

{\bf Hamiltonian formalism:}
It is necessary to transform to $A_0 = 0$ gauge.  I ignore technicalities,
such as fixing the residual degrees of freedom for the $A_i$ fields
\cite{Gervais:1978kn,Goldstone:1978he,Jackiw:1979ur,Christ:1980ku},
as these do not affect my analysis.
Under a gauge transformation, 
\begin{equation}
  A_\mu(\vx,\tau) \rightarrow \frac{1}{-ig} \Omega^\dagger(\vx,\tau) D_\mu  \Omega(\vx,\tau) \; .
  \end{equation}
The gauge transformation which implements $A_0 = 0$ gauge is just
$\Omega(\vx,\tau) = {\bf L}(\vx;\tau,0)$.
Since in general $\Omega(\vx,1/T) \neq \Omega(\vx,0)$, when $A_0(\vx,\tau)=0$
the $A_i(\vx,\tau)$ are no longer periodic in $\tau$.

In the Hamiltonian formalism, the basic variables are the spatial gauge fields, $\vec{A}$, whose conjugate
momenta are the color electric fields, $\vec{E} = \partial_0 \vec{A}$.  For a quark field $\psi$
the conjugate momentum is $\overline{\psi}$ \cite{Gervais:1978kn,Goldstone:1978he,Jackiw:1979ur,Christ:1980ku}.
The Hamiltonian density is
\begin{equation}
  \hamil(\vx) = {\rm tr} \left( \vec{E}^2(\vx) + \vec{B}^2(\vx) \right) \; ;
  \label{eq:hamil}
\end{equation}
as
the test quark only enters into the Lagrangian as $\overline{\psi} \partial_0 \psi$,
it drops out of the Hamiltonian.

To ensure the conservation of color electric charge, however, it is still necessary
to impose Gauss's law.  For this it is convenient to introduce an auxiliary field, $\chi(\vx)$:
\begin{equation}
  \hamil_{\rm Gauss}(\vx) = {\rm tr}
  \left( \chi(\vx) \left(\vec{D} \cdot \vec{E}(\vx)  - g \, \chg(\vx)  \right)\right) \; ,
  \label{eq:cons}
\end{equation}
where  $\chg^a(\vx) = \psi^\dagger(\vx) t^a \psi(\vx)$ is the color charge of the test quark
and the $t^a$ are the generators in the fundamental representation, $a = 1\ldots(N^2-1)$.
Since only particles without spin enter in the effective Lagrangian, it is not necessary to bother with Dirac matrices.
The color charge $\chg$ transforms homogeneously
in the adjoint representation, $\chg(\vx) \rightarrow \Omega^\dagger(\vx) \, \chg(\vx) \, \Omega(\vx)$, as
does the constraint field $\chi(\vx)$
\footnote{On the lattice, $\chi$ lives on sites, not links.}.
For a Polyakov loop, the color charge $\chg(\vx)$ is a single
$\delta$-function in $\vx$; for a Wilson loop, there are two $\delta$-functions, and so on for more loops.

As discussed by Gervais and Sakita \cite{Gervais:1978kn},
states for test quarks must be included in the partition function.
To understand how they contribute, I ask:

{\it How does the exponential of a trace become the trace of an exponential?}

That is, how does the test charge $\chg$ in Eq. (\ref{eq:cons}) transform into the 
Wilson and Polyakov loops of Eqs. (\ref{eq:wilson_loop}) and (\ref{eq:polyakov_loop})?

The answer is an exercise in the character for a representation of a Lie group
\cite{Georgi:2000vve,Cvitanovic:2008zz,Greiner:2001,Hall:2015}.
This was used originally by Susskind \cite{Susskind:1979up}
\footnote{In Eq. (60) of Ref. \cite{Susskind:1979up}, the sum is over the representations of the electric
field, to give the character of $\chi$, as in Eq. \ref{eq:trace_explicit}.  
The product of characters which arise for the two point function of Polyakov loops, Eq. (68), is given
without comment.  I show that this is due to the sum over states of the test charge.},
and is related to an analysis by Greiner and M\"uller
\cite{Greiner:2001}.

A representation $\rep$ of the $SU(N)$ group is characterized by a Young tableaux, which
are $N-1$ integers, $n_1,n_2\ldots n_{N-1}$, where $n_1 \geq n_2 \ldots \geq n_{N-1}$
\cite{Georgi:2000vve,Cvitanovic:2008zz,Greiner:2001,Hall:2015}.  For the case of $SU(2)$, there is only one row, 
where $n_1$ equals the spin, $n_1 = j = 0,1,2\ldots$.

The representations which contribute to the states of the electric field are denoted as $|\rep(\vx)\rangle$,
and that for the test quark as $|\tlrep(\vx)\rangle$.
While the test quark lies in a fixed representation at a few points in space,
all representations contribute to the state space of the electric field 
at each point in space.  For the example of $SU(2)$,
all $j(\vx)$ contribute to the electric field at each $\vx$, while only a single $\widetilde{j}(\vx)$ contributes
to that of the test quark, at the point where the Polyakov loop lies.

The expectation value of the Polyakov loop is given by
\begin{eqnarray}
  \langle \pol(\vy)  \rangle &=& \widetilde{\sum} \; \int {\cal D} \chi(\vx) \; \exp \left(
  - \int d^3 x  \, \hamil_0(\vx)/T \right) \; , \nonumber \\
  \hamil_0(\vx) &=& \hamil(\vx) + {\rm tr} \; \chi(\vx) \, \vec{D} \cdot \vec{E}(\vx)
  + \hamil_\chg(\vx) \; , \nonumber \\
  \hamil_\chg(\vx) &=& - {\rm tr} \; \chi(\vx) \, \chg(\vx) \; .
  \label{eq:constrained_hamiltonian}
\end{eqnarray}
The sum over states, $\widetilde{\sum}$, includes those for the gauge field, the $\vec{A}(\vx)$
and $\vec{E}(\vx)$, {\it and} the states for the test quarks, $\psi(\vx)$ and $\overline{\psi}(\vx)$.
It is also necessary to include a path integral for the constraint field, $\chi(\vx)$.
Dynamical quarks can be included directly.

I note that it is meaningful to compute the expectation
value of a single Polyakov loop in a non-Abelian gauge theory, as the color charge is always screened.
At low temperature this happens either because of confinement (without dynamical quarks),
or the pair production of mesons (with dynamical quarks).  At high temperature, there is always
Debye screening
\footnote{
The Abelian theory is different, as while there is Debye
screening at high temperature, there is no screening at low temperature.
Further, in a finite box the presence of a test charge is inconsistent with
periodic boundary conditions \cite{Hilf:1983ra}, and so
open boundary conditions must be used.  
In the unscreened phase of the Abelian theory it is still possible to measure the potential
between a test charge and anti-charge, ${\cal V}_\infty(\vx)$.}.
Typical of a system with
the spontaneous breaking of a global symmetry, the expectation value of the Polyakov loop
is only well defined after introducing
an appropriate
\footnote{The appropriate sources for Polyakov loops must involve a sum over an infinite number of loops.
This is because the matrix for any representation of $SU(N)$ is traceless, so 
for a single loop the term linear in $A_\mu$ vanishes at small $A_\mu$.
This holds for a sum over any finite number of loops, but fails if the sum is infinite.  
An appropriate source is that for which an infinitesimal source generates an expectation value which is
also infinitesimal. For details, see Refs.
\cite{KorthalsAltes:2019yih,KorthalsAltes:2020ryu,Hidaka:2020vna}.}
external source for the corresponding field, and then tuning that source to zero.

It is easy to perform the sum over states for the Polyakov loop in Eq. \ref{eq:constrained_hamiltonian},
as the only quantum number carried by the test quark is that for color electric charge.
Since the color charge transforms homogeneously under gauge transformations, I can assume
that it is a diagonal matrix.  Then if the quark and anti-quark
have color $1$, $\chg(\vx)^a = t^a_{1 1} \delta^3(\vx-\vy)$, where $\vy$ is the position of the loop, and contributes 
to $\pol(\vx)$ as $\sim \exp(i g \chi^a\, t^a_{1 1}) \delta^3(\vx-\vy)$.
If the quark and anti-quark have color $2$, the contribution to $\pol(\vx)$ is
$\sim \exp(i g \chi^a \, t^a_{2  2})\delta^3(\vx-\vy)$, and
so on.  

{\it Thus the trace over states of the test quark is just a trace over color},
\begin{equation}
  \langle \pol(\vy) \rangle = \sum \int {\cal D} \chi(\vx) \; {\rm e}^{  - \int \, \hamil_{\rm 0}/T} \; 
  {\rm tr} \; {\rm e}^{i g \chi(\vy)} \; .
  \label{eq:trace_explicit}
\end{equation}
Having summed over the states for the test quark and anti-quark, the remaining sum,
$\sum$, is only for the $\vec{A}$ and $\vec{E}$, along with the path integral over the constraint field, $\chi(\vx)$.

From this derivation, it is apparent
that $\chi(\vx)$ corresponds to the time-like component of the gauge potential
in the Lagrangian formalism, $A_0(\vx)$ \cite{Gervais:1978kn,Goldstone:1978he,Jackiw:1979ur,Christ:1980ku}.
The transformation
from $A_0=0$ gauge in the Lagrangian formalism is given by identifying the gauge transformation
$\Omega(\vx,1/T) = {\bf L}(\vx;,1/T,0)$ with $\exp(i g \chi(\vx))$.
After averaging over the spatial volume, this agrees with Eq. (\ref{eq:free_energy}).
Similarly, the thermal Wilson loop $\wilson_T$ equals
\begin{equation}
  \wilson_T = {\rm tr} \; {\rm e}^{i g \chi(\vx) }  \; {\bf L}(\vx;\vec{0})\;
         {\rm e}^{- i g \chi(\vec{0}) }\; {\bf L}(\vec{0}; \vx) \; .
      \label{eq:wilson_thermal_hamil}
\end{equation}

The generalization to higher representations of the test quark
is direct, with the generator is given by a bird track diagram, Eq. (4.35) of Cvitanovic
\cite{Cvitanovic:2008zz}.  The sum over all color states
is given by summing over all of the legs of the bird track.

In the Lagrangian formalism, the holonomous potential for the eigenvalues of the thermal Wilson line
is familiar, and computed by expanding about a constant, background field
$A_0 \neq 0$
\cite{Gross:1980br,Weiss:1980rj,Bhattacharya:1990hk,Belyaev:1991gh,Bhattacharya:1992qb,Gocksch:1993iy,KorthalsAltes:1993ca,KorthalsAltes:1994be,KorthalsAltes:1996xp,KorthalsAltes:1999cp,Giovannangeli:2001bh,Giovannangeli:2002uv,Giovannangeli:2004sg,Dumitru:2013xna,Guo:2014zra,Nishimura:2017crr,Guo:2018scp,KorthalsAltes:2019yih,KorthalsAltes:2020ryu,Hidaka:2020vna,Guo:2020jvc}.
In the Hamiltonian formalism, non-trivial holonomy becomes the factor of $\exp(i g \chi(\vx))$, which enters
as an imaginary chemical potential for the color charge.

In the gauge theory without dynamical quarks, the holonomous potential manifestly exhibits the global
$Z(N)$ degeneracy for the $SU(N)/Z(N)$ theory.  With dynamical quarks, however, depending upon the representation of
the quarks and the color, it is possible to have metastable states with negative pressure
\cite{Belyaev:1991np,Chen:1992sa,Kogan:1993bz,Smilga:1993vb,Smilga:1996cm}.  This occurs because 
the zero of the potential for non-trivial holonomy has an absolute significance, as the pressure with
with vanishing holonomy.  However, while a bubble of such a metastable state can have negative pressure, 
it only lasts as long as it takes the surface of the bubble to
collapse upon itself, decaying through cavitation \cite{Rajagopal:2009yw,Brennan:2014}.

{\bf 't Hooft loops:}
Wilson loops represent the propagation of test electric charge.  't Hooft first constructed a dual order
parameter to the Wilson loop, which represents the propagation of a test magnetic charge
\cite{tHooft:1979rtg,tHooft:1980kjq}.  The Wilson and 't Hooft loops satisfy a commutation law, which in vacuum
excludes the simultaneous confinement of electric and magnetic charges.  The
commutation law follows by considering the Wilson
loop as the propagator for a test electric charge: as a tiny Wilson loop encircles a 't Hooft loop,
by definition the phase of a test charge (in the fundamental representation)
changes by ${\rm e}^{2 \pi i/N}$.

At nonzero temperature in the $SU(N)/Z(N)$ gauge theory, due to the global $Z(N)$ symmetry there are $N$ degenerate
vacua in the deconfined phase.  At high temperature, it is possible to consider
a box which is long in one spatial direction, and to compute the interface
tension between a $Z(N)$ vacuum at one end of the box, and a different $Z(N)$ vacuum at the other
This interface tension can be computed semi-classically \cite{Bhattacharya:1990hk,Belyaev:1991gh,Bhattacharya:1992qb}
from the holonomous potential
\cite{Gocksch:1993iy,KorthalsAltes:1993ca,KorthalsAltes:1994be,KorthalsAltes:1996xp,KorthalsAltes:1999cp,Giovannangeli:2001bh,Giovannangeli:2002uv,Giovannangeli:2004sg,Dumitru:2013xna,Guo:2014zra,Nishimura:2017crr,Guo:2018scp,KorthalsAltes:2019yih,KorthalsAltes:2020ryu,Hidaka:2020vna,Guo:2020jvc}.
Korthals-Altes, Kovner, and Stephanov showed that the interface tension
is equivalent to an area law for the spatial 't Hooft loop
\cite{Korthals-Altes:1999cqo,Korthals-Altes:2000tia}.
Numerical simulations on the lattice have measured
how the 't Hooft loop changes with temperature 
\cite{deForcrand:2000fi,deForcrand:2001nd,deForcrand:2005pb}.  See, also, Refs. \cite{Reinhardt:2002mb,Reinhardt:2007wh}.

In the Hamiltonian form, a domain wall can be constructed by using a constant $\chi$ field, corresponding
to constant $A_0$ in the Lagrangian formalism.
The simplest model to study is the Abelian theory in $1+1$ dimensions, where the object analogous to a domain
wall in higher dimensions is a soliton.
Since gauge fields have no physical degrees of freedom in two spacetime dimensions,
it is necessary to add dynamical fermions.
Adding massless fields gives the Schwinger model, but this theory behaves contrary to naive
expectation, as even fractional test charges are screened by dynamical fields
with integral charge \cite{Coleman:1975pw,Gross:1995bp,Dempsey:2021xpf}.

If the dynamical fields are massive, though, dynamical fermions with
integral charge do not screen fractional test charges \cite{Coleman:1975pw,Gross:1995bp,Dempsey:2021xpf}.
Smilga first noted the existence of thermal solitons in the massive Schwinger model \cite{Smilga:1993vb}.
In the Euclidean Lagrangian, one takes a background, classical
field $A_0^{\rm cl} = 2 \pi T\, q/e$, where $e$ is the Abelian coupling constant;
in the Hamiltonian form, one takes a similar background for the constraint
field, $\chi$.  As $q$ is a periodic variable, a thermal soliton interpolates from
$q = 0$ at $x = -\infty$ to $q=1$ at $x=+\infty$.
At high temperature, $T \gg m$, the potential for
$q$ generated at one loop order is
$\sim T^2$ times a periodic function of $q$, Eq. (3.9) of Ref. \cite{Smilga:1993vb}.
At low temperature, the potential is Boltzmann suppresed,
$\sim {\rm e}^{-m/T}$, and vanishes smoothly as $T \rightarrow 0$.

I suggest that such solitons are stable.
At $T\neq 0$ imaginary time is topologically equivalent to a torus, $S^1$.
As $q$ is a periodic variable, then, mappings from space onto $q$ are determined by the first homotopy group,
$\pi_1(S^1) = Z$, which is the set of the integers.

Smilga and others 
argued that thermal solitons are unphysical
\cite{Smilga:1993vb,Hansson:1994ep,Kiskis:1995iq,Smilga:1996cm},
%
%
I suggest that they represent new, collective excitations at $T \neq 0$, which evaporate 
smoothly as $T \rightarrow 0$.  This can be studied numerically 
at nonzero temperature in real time, using either tensor networks on a classical computer
\cite{Banuls:2009xxx,Banuls:2019qrq,Lerose:2022sxm,Frias-Perez:2022wwa},
or even with the noisy intermediate-scale quantum computers which are available at present.  This is similar
to studying the screening of background electric fields at nonzero $\theta$
\cite{Klco:2018kyo,Chakraborty:2020uhf,Kharzeev:2020kgc,Pla:2020tpq,Shaw:2020udc,deJong:2021wsd,Florio:2021xvj,Honda:2021aum,Honda:2021ovk,Pederiva:2021tcd}.

There are many other problems which can be addressed with the formalism developed here.
In particular, deep inelastic scattering is usually described by the propagation of time-like Wilson lines
\cite{Echevarria:2020wct}. My approach can be adapted to the light-front directly
\cite{,Lerose:2022sxm,Frias-Perez:2022wwa,Wilson:1994fk},
especially using quantum computers \cite{Kreshchuk:2020dla,Kreshchuk:2020kcz,Kreshchuk:2020aiq}.

Lastly, if thermal solitons are stable in $1+1$ dimensions, presumably thermal domain walls exist in
$3+1$ dimensions.  In the early universe, they can arise
from the $U(1)$ of electromagnetism, when regions which are causally disconnected from one another
first come in contact.  
They persist until the thermal potential for the domain wall of the
lightest electrically charged particle, which is the
electron, is Boltzmann suppressed.
As this temperature is presumably below that for nucleosynthesis, and as 
domain walls dominate the energy density of the universe,
such thermal $U(1)$ domain walls could be of cosmological consequence.

\acknowledgments
This work was inspired by a talk which Tom Cohen gave 
virtually at the Yukawa Institute for Theoretical Physics, during
a workshop on the “QCD phase diagram and lattice QCD”, YITP-W-21-09, October 25-29, 2021.
I thank him, M. Creutz, A. Florio, L. Glozman, C. Korthals-Altes, S. Mukherjee, P. Petreczky, O. Philipsen, E. Poppitz,
and A. Smilga for discussions.
This research was supported by the U.S. Department of Energy under contract DE-SC0012704.
I was also led to consider this problem because of the support of
the U.S. Department of Energy, Office of Science,
National Quantum Information Science Research Centers, Co-design Center for Quantum Advantage (C$^2$QA)
under contract number DE-SC0012704.  

%

\end{document}